\documentclass[twocolumn,english,aps,showpacs,superscriptaddress]{revtex4-1}
\usepackage[latin9]{inputenc}
\usepackage{mathtools}
\usepackage{amsmath}
\usepackage{amssymb}
\usepackage{cancel}
\usepackage{mathdots}
\usepackage{stmaryrd}
\usepackage{stackrel}
\usepackage{graphicx}
\PassOptionsToPackage{version=3}{mhchem}
\usepackage{mhchem}
\usepackage{wasysym}
\usepackage{esint}

\makeatletter

\newcommand*\LyXThinSpace{\,\hspace{0pt}}

\usepackage{verbatim}

\usepackage{bm}
\usepackage{xcolor}
\usepackage{subfigure}

\providecommand{\tabularnewline}{\\}

\@ifundefined{textcolor}{}{%
 \definecolor{BLACK}{gray}{0}
 \definecolor{WHITE}{gray}{1}
 \definecolor{RED}{rgb}{1,0,0}
 \definecolor{GREEN}{rgb}{0,1,0}
 \definecolor{BLUE}{rgb}{0,0,1}
 \definecolor{CYAN}{cmyk}{1,0,0,0}
 \definecolor{MAGENTA}{cmyk}{0,1,0,0}
 \definecolor{YELLOW}{cmyk}{0,0,1,0}
}

\usepackage{bm}

\newcommand{\beq}{\begin{equation}}
\newcommand{\eeq}{\end{equation}}
\newcommand{\bea}{\begin{eqnarray}}
\newcommand{\eea}{\end{eqnarray}}

\newcommand{\fvec}[1]{\boldsymbol{#1}}

\newcommand{\half}{\frac{1}{2}}

\newcommand{\rmd}{{\rm d}}

\usepackage{babel}

\usepackage{babel}

\makeatother

\usepackage{babel}
\begin{document}

\title{Time-reversal symmetry-breaking nematic superconductivity in FeSe}

\author{Jian Kang}

\affiliation{National High Magnetic Field Laboratory, Florida State University,
Tallahassee, Florida, 32304, USA}
\email{jian.kang@fsu.edu}

\selectlanguage{english}%

\author{Andrey V. Chubukov}

\affiliation{School of Physics and Astronomy, University of Minnesota, Minneapolis
55455, USA}

\author{Rafael M. Fernandes}

\affiliation{School of Physics and Astronomy, University of Minnesota, Minneapolis
55455, USA}
\begin{abstract}
FeSe is a unique member of the family of iron-based superconductors,
not only because of the high values of $T_{c}$ in FeSe monolayer,
but also because in bulk FeSe superconductivity emerges inside a nematic
phase without competing with long-range magnetic order. Near $T_{c}$,
superconducting order necessarily has $s+d$ symmetry, because nematic
order couples linearly the $s$-wave and $d$-wave harmonics of the
superconducting order parameter. Here we argue that the near-degeneracy
between $s$-wave and $d$-wave pairing instabilities in FeSe, combined
with the sign-change of the nematic order parameter between hole and
electron pockets, allows the superconducting order to break time-reversal
symmetry at a temperature $T^{*}<T_{c}$. The transition from an $s+d$
state to an $s+\mathrm{e}^{i\alpha}d$ state should give rise to a
peak in the specific heat and to the emergence of a soft collective
mode that can be potentially detected by Raman spectroscopy.
\end{abstract}
\maketitle

\section{Introduction}

The discovery of FeSe brought renewed interest in the field of unconventional
superconductors, not only because FeSe-based compounds display the
highest superconducting (SC) transition temperatures among all iron-superconductors,
but also because of their unique phase diagram \cite{recent}. Indeed,
in contrast to most Fe-based compounds, bulk FeSe undergoes nematic
and superconducting transitions without displaying long-range antiferromagnetic
order \cite{Baek15,Bohmer15}. The microscopic origin of this unusual
behavior has been the subject of intense debates~\cite{RGFeSe,CDWFeSe,SiFeSe,NevidomskyyFeSe,KivelsonFeSe,Wang16,Glasbrenner15}.
Regardless of the microscopic origin of nematicity, the phase diagrams
of pure and doped FeSe provide a remarkable opportunity to investigate
the interplay between nematicity and superconductivity without the
interfering effects of the antiferromagnetic order observed near the
onset of nematicity in other iron-based materials \cite{Shibauchi17,Canfield17}.

It is well established that nematic and superconducting orders coexist
microscopically in FeSe, with the former onsetting at $T_{s}\approx90$
K~\cite{MKWu,CavaFeSe} and the latter at $T_{c}\approx8$ K~\cite{MKWu}.
Recent experimental~\cite{Dong16,shin,Zhou18,Borisenko18,Rhodes18,BrianSTM17,Wirth17,Taillefer16,Meingast17,Matsuda,hanaguri}
and theoretical works~\cite{Mukherjee15,Kreisel17,Kang18,Fanfarillo18}
have highlighted how the modifications in the orbital compositions
of the Fermi surfaces below the nematic transition influence the superconducting
gap structure, and particularly the gap anisotropy on both hole and
electron pockets. These gap anisotropies have been observed directly
by ARPES~\cite{Dong16,shin,Zhou18,Borisenko18,Rhodes18} and STM~\cite{BrianSTM17,Wirth17,hanaguri},
and also indirectly in specific heat and thermal conductivity measurements~\cite{Taillefer16,Meingast17,Matsuda}.

General models for the pairing interaction in Fe-based SC have revealed
closely competing $s^{+-}$ and $d$-wave parring channels, with the
latter even winning over the former in certain models \cite{Zhang09,LAHA,Stanev10,Hanke12,Khodas12,Fernandes_Millis13,Millis13,Varelogiannis15,Kang14}.
However, in Fe-pnictides such near-degeneracy holds only far enough
from the magnetically ordered phase, otherwise the $(\pi,0)$/$(0,\pi)$
stripe-type magnetic fluctuations favor $s^{+-}$ pairing \cite{Maier18}.
In FeSe the situation is different. First, there is no magnetic order.
If one takes this as evidence that magnetic fluctuations are not strong
and treats the pairing within the Kohn-Luttinger scenario, one finds
(see below) that $s-$wave and $d-$wave pairing amplitudes are quite
comparable. Second, if one takes a different point of view and assumes
that magnetic fluctuations in FeSe are strong enough to justify a
spin-fluctuation approach, one still has to include into consideration
not only $(\pi,0)$/$(0,\pi)$ fluctuations, but also $\left(\pi,\pi\right)$
Néel-type magnetic fluctuations \cite{DHLee16,EAKim17}, as both have
been observed in neutron scattering~\cite{Zhao16}. The stripe magnetic
fluctuations enhance the pairing strength in the $s^{+-}$ pairing
channel, and the $\left(\pi,\pi\right)$ fluctuations do the same
in the $d$-wave channel \cite{Fernandes_Millis13}. The existence
of both fluctuations again keeps the $s^{+-}$ and $d$-wave pairing
amplitudes comparable.

\begin{figure}[htbp]
\centering{}\includegraphics[width=0.9\columnwidth]{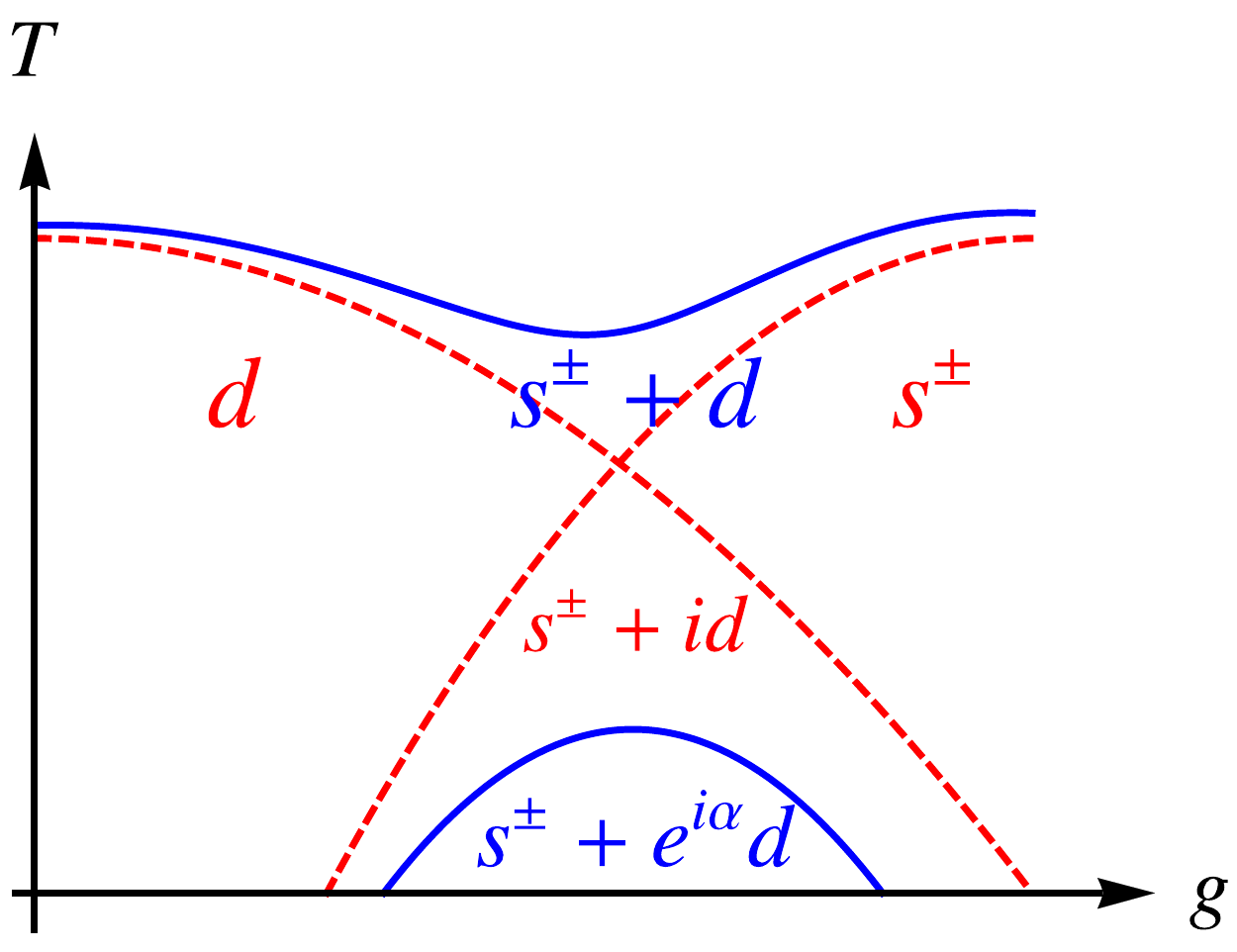}
\caption{Schematic figure summarizing our main results. As function of a tuning
parameter $g$ (which in our paper is the ratio between intra-orbital
and inter-orbital inter-band pairing interactions), the superconducting
state changes from $d$-wave to $s^{+-}$-wave in the tetragonal phase,
giving rise to an $s+id$ state near the degeneracy point (dashed
red curves). In the nematic phase, the superconducting state becomes
$s+d$ near $T_{c}$, which is enhanced near the $s$-wave/$d$-wave
degeneracy point. At low temperatures, an $s+\mathrm{e}^{i\alpha}d$
state can be stabilized (solid blue curves). Such an exotic state,
which breaks both time-reversal and tetragonal symmetries, is much
more favored for a sign-changing nematic state, as compared to a sign-preserving
nematic state.}
\label{Fig:schematic}
\end{figure}

It is well known that proximate $s^{+-}$ and $d$-wave states can
lead to the emergence at low enough $T$ of an exotic superconducting
state that breaks time-reversal symmetry (TRS): the $s+id$ state.
It emerges as the lowest-energy state because it gaps out all states
on the Fermi surfaces and by this maximizes the gain of the condensation
energy. An $s+id$ state has been proposed to exist in strongly hole-doped
and strongly electron-doped Fe-pnictides~\cite{Zhang09,Stanev10,Hanke12,Khodas12,Fernandes_Millis13},
but it has not been yet unambiguously detected in experiments. As
we just said, in FeSe $s^{+-}$ and $d$-wave states are likely closer
than in Fe-pnictides \cite{Mukherjee15}, so FeSe seems a natural
candidate to search for $s+id$ order. However, there is a caveat
\textendash{} the nematic order couples linearly the $s-$wave and
$d-$wave channels. Because of this coupling, the superconducting
order parameter near $T_{c}$ necessarily has $s+d$ symmetry rather
than $s+id$ \cite{Millis13,Varelogiannis15,Kang14}. An $s+d$ superconducting
order preserves time-reversal symmetry and just changes the anisotropy
of the gap function. If the linear coupling between $s-$wave and
$d-$wave order parameters is strong enough, $s+d$ state persists
down to $T=0$. If, however, it is weak, the system may undergo
a transition at some $T<T_{c}$ into a time-reversal symmetry breaking
(TRSB) state (see Figure \ref{Fig:schematic}).

In general, the gap function of a superconducting state with $s$-wave
and $d$-wave components is parametrized by

\begin{equation}
\Delta=\Delta_{s}+\mathrm{e}^{i\alpha}\Delta_{d}\label{1}
\end{equation}
For definiteness, we assume that both $\Delta_{s}$ and $\Delta_{d}$
are real. The relevant parameter in (\ref{1}) is the relative phase
$0\leq\alpha\leq\pi$ between the two order parameters. In the $s+d$
state, $\alpha=0$ or $\pi$, wheeras in $s+id$ state $\alpha=\pm\pi/2$.
Other values of $\alpha$ describe nematic superconducting states
which also break TRS.

In this work we analyze the gap structure of FeSe below $T_{c}$ by
solving the set of non-linear gap equations on the different hole
and electron pockets. We take as input the fact that in FeSe the nematic
order parameter changes sign between hole and electron pockets \cite{Shimojima15,Kontani16,RGFeSe,Rhodes18}.
We argue that for a sign-changing nematic order, the linear coupling
between $s-$wave and $d-$wave gap components is much smaller than
it would be if nematic order was sign-preserving. We analyze the gap
structure and show that for parameters appropriate for FeSe it is
quite likely that below some $T^{*}<T_{c}$, $\alpha$ becomes different
than $0$ or $\pi$, i.e., the system undergoes a transition into
TRSB state.

Such a SC-to-SC transition is manifested by the softening of the collective
mode associated with the fluctuations of $\alpha$. This can be probed
by Raman spectroscopy. The signatures of the transition into the TRSB
state can be also found by measuring thermodynamic quantities, such
as the specific heat. Interestingly, recent specific heat measurements
on FeSe have reported a peak well below $T_{c}$~\cite{HHWen17}.
We conjecture that this feature could be due to the formation of the
TRSB state.

The paper is organized as follows. In Sec. II we introduce our microscopic
model with hole and electron pockets and onsite Hund and Hubbard
interactions. We obtain the effective pairing interactions in the
$s^{+-}$ and $d-$wave channels within the Kohn-Luttinger formalism
and rationalize an effective model with $s-$wave and $d-$wave attractive
interactions of comparable magnitudes. In Sec. III we study the pairing
in the tetragonal phase and obtain the TRSB $s+id$ state for some
range of system parameters. In Sec. IV we analyze the pairing in the
presence of nematic order. We show that near $T_{c}$ the pairing
symmetry is necessarily $s+d$, but TRSB state may still emerge at
a lower $T$. In this section we also compare the effects of sign-changing
and sign-preserving nematic order. We argue that TRSB state is substantially
more likely when the nematic order is sign-changing, as in FeSe. In
Sec. V we discuss experimental signatures of the transition into the
TRSB state, in particular the softening of the collective mode associated
with the fluctuations of the relative phase between $s-$wave
and $d-$wave order parameters. Section VI presents our conclusions. 
Appendices A and B contain additional details not discussed in
the main text,

\section{Microscopic model}

\subsection{Non-interacting terms}

Our microscopic Hamiltonian contains a non-interacting part, $\mathcal{H}_{0}$,
and an interacting part, $\mathcal{H}_{\mathrm{int}}$. The former
is constructed based on ARPES measurements, which find, above the
nematic transition temperature $T_{s}$, two small hole pockets at
the center of the Brillouin zone and two small electron pockets centered
at $\mathbf{Q}_{X}=\left(\pi,0\right)$ and $\mathbf{Q}_{Y}=\left(0,\pi\right)$
in the Fe-only Brillouin zone \cite{Coldea15,Borisenko16,Coldea17,Borisenko18}.
The spectral weight of excitations near the hole pockets comes predominantly
from the $d_{xz}$ and $d_{yz}$ Fe orbitals. As a result, the hole-band
operators $h_{1,\mathbf{k}}$ and $h_{2,\mathbf{k}}$ can be expressed
in terms of the orbital operators $d_{xz,\mathbf{k}}$ and $d_{yz,\mathbf{k}}$
as \cite{Vafek13,Fernandes17}:

\begin{align}
h_{1,\mathbf{k}} & =i\left(d_{xz,\mathbf{k}}\cos\theta_{\mathbf{k}}+d_{yz,\mathbf{k}}\sin\theta_{\mathbf{k}}\right)\nonumber \\
h_{2,\mathbf{k}} & =i\left(-d_{xz,\mathbf{k}}\sin\theta_{\mathbf{k}}+d_{yz,\mathbf{k}}\cos\theta_{\mathbf{k}}\right)\label{hole_operators}
\end{align}

The imaginary pre-factors are introduced for convenience, as later
we will search for TRSB solutions of the gap equations. The relationship
between the Bogoliubov parameter $\theta_{\mathbf{k}}$ and the polar
angle $\theta$ around the hole pockets (measured with respect to
$k_{x}$) depends on the tight-binding parameters. For our purposes,
it is sufficient to consider the special case of circular hole pockets,
in which case $\theta_{\mathbf{k}}=\theta$. Because the hole pockets
are small, we can approximate their dispersions as parabolic, $\varepsilon_{h_{i}}=\mu_{h}-k^{2}/2m_{i}$,
with 
$m_{2}>m_{1}$.

The electron pockets are predominantly formed out of $d_{xz/yz}$
and $d_{xy}$ orbitals. We describe them by the operators $e_{X/Y,\mathbf{k}}$.
We use as input the results of earlier renormalization group (RG)
studies~\cite{Classen17} that the interactions involving fermions
from $d_{xy}$ orbitals flow to smaller values than the ones involving
fermions from $d_{xz}$ and $d_{yz}$ orbitals. The smallness of the
interactions involving $d_{xy}$ orbitals has also been proposed in
strong-coupling approaches \cite{Kotliar11,Medici14,Fanfarillo17}
and phenomenologically in recent studies of the linearized gap equation
in FeSe~\cite{BrianSTM17,Kang18}. To simplify the analysis,
we then neglect fermions from the $d_{xy}$ orbital in the pairing
problem and approximate the excitations near the $X$ pocket as $d_{yz}$
($e_{X,\mathbf{k}}=d_{yz,\mathbf{k}+\mathbf{Q}_{X}}$) and near the
$Y$ pocket as $d_{xz}$ ($e_{Y,\mathbf{k}}=d_{xz,\mathbf{k}+\mathbf{Q}_{Y}}$),
with electron-band dispersions $\varepsilon_{X,Y}=-\mu_{e}+k_{x}^{2}/\left(2m_{X,Y}\right)+k_{y}^{2}/\left(2m_{Y,X}\right)$.
This approximation substantially simplifies the analysis of the transition
into the TRSB state at $T^{*}$. The inclusion of $d_{xy}$ orbitals
does not change the main results, as it only shifts the value of $T^{*}$.
By the same reason, we also neglect spin-orbit coupling \cite{Fernandes_Vafek}
and the variation of the size of the hole pockets along the $k_{z}$
direction. The $k_{z}$ variation is relevant for the understanding
of the orbital composition of the hole pockets in the nematic phase
and of the gap anisotropy in the $s+d$ phase~\cite{Borisenko18,Rhodes18,Kang18},
but does not qualitatively alter the physics of the transition into
the TRSB state.

\subsection{Interaction terms}

The interacting part of the Hamiltonian $\mathcal{H}_{\mathrm{int}}$
is responsible for the SC instability. Let us first analyze SC in
the absence of nematic order. As discussed in the introduction, there
are two approaches to the pairing instability. One is to start with
bare interactions, such as the onsite Hund and Hubbard interactions,
and analyze the pairing to second order in perturbation theory (the
Kohn-Luttinger approach). Another is to adopt the semi-phenomenological
spin-fluctuation scenario and analyze the pairing mediated by spin
fluctuations.

Within the first scenario, both $s-$wave and $d-$wave components
of the interaction emerge when one converts from orbital to band basis
using (\ref{hole_operators}). In the tetragonal phase the effective
BCS Hamiltonian factorizes between $s-$ and $d-$channels. Symmetry
analysis shows~\cite{RGFeSe,Vafek13} there are three relevant pairing
interactions: one between fermions on the two hole pockets, another
between fermions on the two electron pockets, and the third between
fermions on hole and on electron pockets. To be consistent with the
notation in earlier works we label these three interactions respectively
as $U_{4}^{s(d)},U_{5}^{s(d)}$ and $U_{3}^{s(d)}$.

To write the BCS Hamiltonian in a compact form we follow~\cite{RGFeSe}
and introduce the pair operators
\begin{align}
\kappa_{\mu\mu'}^{e}=e_{\mu\uparrow}e_{\mu'\downarrow}\,,\quad\kappa_{\mu\mu'}^{h}=h_{\mu\uparrow}h_{\mu'\downarrow}\,.\label{Co}
\end{align}
where $\mu=1,2$, $e_{1}=e_{Y}$ and $e_{2}=e_{X}$, and
\begin{align}
\kappa_{s}^{e(h)} & =\kappa_{11}^{e(h)}+\kappa_{22}^{e(h)}\nonumber \\
\kappa_{d}^{e(h)} & =\kappa_{11}^{e(h)}-\kappa_{22}^{e(h)}
\end{align}

The pairing Hamiltonian is then given by:
\begin{align}
H_{\kappa}=H_{\kappa_{s}}+H_{\kappa_{d}},
\end{align}
where
\begin{align}
H_{\kappa_{s}} & =U_{5}^{s}(\kappa_{s}^{e})^{\dag}\kappa_{s}^{e}+U_{4}^{s}(\kappa_{s}^{h})^{\dag}\kappa_{s}^{h}+U_{3}^{s}\left((\kappa_{s}^{e})^{\dag}\kappa_{s}^{h}+h.c.\right)\label{aa_1}\\
H_{\kappa_{d}} & =U_{5}^{d}(\kappa_{d}^{e})^{\dag}\kappa_{d}^{e}+U_{4}^{d}(\kappa_{d}^{h})^{\dag}\kappa_{d}^{h}+U_{3}^{d}\left((\kappa_{d}^{e})^{\dag}\kappa_{d}^{h}+h.c.\right)\label{aa_1_1}
\end{align}
We assume momentarily that the densities of states on all pockets
are equal to $N_{F}$. Introducing $u_{i}^{s(d)}=U_{i}^{s(d)}N_{F}/2$
and solving the BCS gap equations in $s-$wave and $d-$wave channels,
we obtain two dimensionless couplings in each channel, corresponding
to same-sign (denoted by $s^{++}$ and $d^{++}$) or opposite-sign
(denoted by $s^{+-}$ and $d^{+-}$) gaps on electron and hole pockets.
One of the two is repulsive for positive $U_{i}^{s(d)}$, whereas
the other can be of either sign, depending on the interplay between
the $U_{5},U_{4}$, and $U_{3}$ interactions. These couplings are
\begin{align}
\lambda^{s} & =\frac{(u_{3}^{s})^{2}-u_{4}^{s}u_{5}^{s}}{u_{4}^{s}+u_{5}^{s}}\label{2}\\
\lambda^{d} & =\frac{(u_{3}^{d})^{2}-u_{4}^{d}u_{5}^{d}}{u_{4}^{d}+2u_{5}^{d}}\label{3}
\end{align}
Note that a positive $\lambda$ implies attraction. Symmetry analysis
shows~\cite{Vafek17} that $\lambda^{s}$ corresponds to the $s^{+-}$
channel and $\lambda_{d}$ to the $d^{++}$ chanel.

The bare values of the interactions $U_{i}^{s(d)}$ are $U_{5}^{s}=U_{4}^{s}=U_{3}^{s}=(U+J)/2,U_{5}^{d}=U_{4}^{d}=U_{3}^{d}=(U-J)/2$.
Substituting into (\ref{2}) and (\ref{3}) we see that the couplings
in both $s$ and $d$ channels vanish. A non-zero $\lambda^{s(d)}$
emerge when we include the renormalizations of the interactions between
given fermions due to the presence of other fermions. If the system
does not show a strong tendency towards a density-wave order, these
renormalizations can be computed to second order in $u_{i}^{s(d)}$.
Still, in systems with hole and electron pockets, some renormalizations
are logarithmically singular, i.e., they depend on $L=\log{W/E}$,
where $W$ is the bandwidth and $E$ is the energy at which we probe
$\lambda^{s}$ and $\lambda^{d}$. Keeping only the logarithmical
terms, and extracting the renormalizations to order $(u_{i}^{s(d)})^{2}$
from the RG equations for the flow of the couplings~\cite{RGFeSe},
we obtain, in terms of $U$ and $J$,
\begin{align}
\lambda^{s} & =4U^{2}N_{F}^{2}\left(1+J/U-2J^{2}/U^{2}\right)L\label{4}\\
\lambda^{d} & =\frac{8}{3}U^{2}N_{F}^{2}(1-J/U+2J^{2}/U^{2})L\label{5}
\end{align}
We see that for $U>J$, both $\lambda_{s}$ and $\lambda_{d}$ are
positive, i.e., the dressed couplings are attractive in both $s-$wave
and $d-$wave channels. For large $U/J$, $\lambda^{s}>\lambda^{d}$,
i.e. the $s-$wave channel is more attractive. However, for $0.65<J/U<1$,
$\lambda^{d}>\lambda^{s}$, implying that the $d-$wave channel is
more attractive. For non-equal densities of states on different pockets
the formulas are more complex, but the key result is the same: the
couplings $\lambda^{s}$ and $\lambda^{d}$ in $s^{+-}$ and $d^{++}$
channels vanish if we use bare interactions but become positive (attractive)
when we include the renormalizations of the interactions to order
$(u_{i}^{s(d)})^{2}$.

From physics perspective, $\lambda^{s(d)}$ becomes attractive because the corrections
to order $(u_{i}^{s(d)})^{2}$ increase the inter-pocket interaction
$u_{3}$ compared to the interactions between fermions near only hole
or only electron pockets. This generally moves the system towards
a stripe magnetic order. For large enough $U/J$, this predominantly
increases the interaction in the $s^{+-}$ channel, but when $U$
and $J$ are comparable, this may increase even more the pairing interaction
in the $d-$wave channel. Note that this is entirely due to the enhancement
of the interaction at the stripe wave-vectors $(0,\pi)/(\pi,0)$.
The $(\pi,\pi)$ interaction between the electron pockets is present
as the inter-pocket component of the interaction $U_{5}^{s(d)}$ ,
but to logarithmical accuracy it is not enhanced compared to the intra-pocket
component of $U_{5}^{s(d)}$.

An alternative to the Kohn-Luttinger approach is the
phenomenological spin-fluctuation approach. Here, one takes as input
the fact that stripe and Neel magnetic fluctuations are enhanced
and considers only inter-pocket interactions with momentum transfer
$(\pi,0)/(0,\pi)$ and $(\pi,\pi)$. In the absence of competing intra-pocket
interactions, $\lambda^{c}$ and $\lambda^{d}$ are definitely positive.
When $U/J$ is large, $(\pi,0)/(0,\pi)$ fluctuations favor $s^{+-}$
pairing. However, $(\pi,\pi)$ fluctuations favor a state with sign-changing gaps
between the electron pockets, which by symmetry is $d-$wave. The
interplay between $\lambda^{s}$ and $\lambda^{d}$ is then determined
by the details of spin-fluctuations near $(\pi,0)/(0,\pi)$ and $(\pi,\pi)$
\cite{Fernandes_Millis13,Millis13}.

We see that in both Kohn-Luttinger and spin-fluctuation approaches,
the couplings $\lambda^{s}$ and $\lambda^{d}$ are attractive, and
their ratio depends on microscopic details. Hereafter we adopt a combined
approach in which we take the elements of both Kohn-Luttinger and
spin-fluctuation treatments. Specifically, we keep in the interaction
Hamiltonian one intra-orbital, inter-band interaction, $V$, and two
inter-orbital, inter-band interactions, $W_{1}$ and $W_{2}$. The
interaction Hamiltonian, projected onto the pairing channel, is
\begin{align}
\mathcal{H}_{\mathrm{SC}} & =V\sum_{\mathbf{k},\mu}d_{\mu,\mathbf{k}\uparrow}^{\dagger}d_{\mu,-\mathbf{k}\downarrow}^{\dagger}d_{\mu,\mathbf{k}'+\mathbf{Q}_{\mu}\downarrow}^{\phantom{\dagger}}d_{\mu,-\mathbf{k}'+\mathbf{Q}_{\mu}\uparrow}^{\phantom{\dagger}}\label{H_SC}\\
 & +W_{1}\sum_{\mathbf{k},\mu\neq\nu}d_{\mu,\mathbf{k}\uparrow}^{\dagger}d_{\mu,-\mathbf{k}\downarrow}^{\dagger}d_{\nu,\mathbf{k}'+\mathbf{Q}_{\nu}\downarrow}^{\phantom{\dagger}}d_{\nu,-\mathbf{k}'+\mathbf{Q}_{\nu}\uparrow}^{\phantom{\dagger}}\nonumber \\
 & +W_{2}\sum_{\mathbf{k},\mu\neq\nu}d_{\mu,\mathbf{k}+\mathbf{Q}_{\mu}\uparrow}^{\dagger}d_{\mu,-\mathbf{k}+\mathbf{Q}_{\mu}\downarrow}^{\dagger}d_{\nu,\mathbf{k}'+\mathbf{Q}_{\nu}\downarrow}^{\phantom{\dagger}}d_{\nu,-\mathbf{k}'+\mathbf{Q}_{\nu}\uparrow}^{\phantom{\dagger}}\nonumber
\end{align}
where $\mu=xz,\,yz$ , $Q_{\mu}=\left(\pi,0\right)$ for $\mu=yz$
and $Q_{\mu}=\left(0,\pi\right)$ for $\mu=xz$. The terms $V$ and
$W_{1}$ are interactions with momentum transfer $Q_{\mu}$, while
$W_{2}$ term describes interaction with momentum transfer $(\pi,\pi)$.

Eq. (\ref{H_SC}) together with the kinetic energy term describe the
pairing in the tetragonal phase. In the nematic phase below $T_{s}$,
two new effects emerge. First, the kinetic energy changes because
nematicity (regardless of its origin) breaks $C_{4}$ lattice rotational
symmetry and gives rise to orbital order, which distinguishes between
$d_{xz}$ and $d_{yz}$ orbitals. The corresponding order parameter
is $\Phi\left(\mathbf{k}\right)=\left\langle n_{xz,\mathbf{k}}\right\rangle -\left\langle n_{yz,\mathbf{k}}\right\rangle $,
where $n_{i}$ is the occupation number operator. This order parameter has two components:
one on hole pockets, $\Phi_{h}\equiv\Phi\left(\mathbf{k}=0\right)$,
another on electron pockets, $\Phi_{e}\equiv\Phi\left(\mathbf{Q}_{X}\right)=-\Phi\left(\mathbf{Q}_{Y}\right)$.
The orbital order with $\Phi_{h}$ and $\Phi_{e}$ adds an additional
term to the kinetic energy in the form
\begin{align}
 & \mathcal{H}_{\mathrm{nem}}=\sum_{\mathbf{k}\sigma}\Phi_{h}\left(d_{xz,\mathbf{k}\sigma}^{\dagger}d_{xz,\mathbf{k}\sigma}^{\phantom{\dagger}}-d_{yz,\mathbf{k}\sigma}^{\dagger}d_{yz,\mathbf{k}\sigma}^{\phantom{\dagger}}\right)+\nonumber \\
 & \sum_{\mathbf{k}\sigma}\Phi_{e}\left(d_{xz,\mathbf{k}+\mathbf{Q}_{Y}\sigma}^{\dagger}d_{xz,\mathbf{k}+\mathbf{Q}_{Y}\sigma}^{\phantom{\dagger}}-d_{yz,\mathbf{k}+\mathbf{Q}_{X}\sigma}^{\dagger}d_{yz,\mathbf{k}+\mathbf{Q}_{X}\sigma}^{\phantom{\dagger}}\right)\label{H_nem}
\end{align}
Orbital order modifies the shapes of the hole and electron pockets:
the hole pockets become elliptical, and one of the electron pockets
becomes peanut-like shaped. This has been observed in ARPES and STM
experiments~\cite{Dong16,shin,Zhou18,Borisenko18,Rhodes18,BrianSTM17,Wirth17,hanaguri}.
The observed geometry of the pockets is reproduced if $\Phi_{h}$
and $\Phi_{e}$ have opposite signs \cite{Shimojima15,Rhodes18}.
In common terminology, such an order is called sign-changing nematic
order. Orbital order also modifies the $s-$wave and $d-$wave components
of the pairing interaction once one converts it from orbital to band
basis, because in the presence of (\ref{H_nem}) the Bogoliubov parameter
$\theta_{k}$ in (\ref{hole_operators}) no longer coincides with
the polar angle around the hole pocket \cite{Kang14}.

The second effect of nematicity is the splitting of the pairing interaction
between $xz$ and $yz$ orbitals already in the orbital basis, i.e.,
in (\ref{H_SC}). In some earlier works this effect has been included
either phenomenologically~\cite{BrianSTM17,Kreisel17}, or semi-phenomenologically,
by invoking spin-nematic scenario and assuming stronger spin fluctuations
at $(\pi,0)$ in the nematic phase~\cite{Millis13,Fanfarillo16,Fanfarillo18,Rhodes18}.
In our study we neglect this effect on the grounds that (i) in the
band basis (which we will use to study pairing) its result is qualitatively
similar to nematicity-induced change of the Bogoliubov parameter $\theta_{k}$;
and (ii) the strength of the $d_{xz}/d_{yz}$ splitting of the pairing
interaction in the orbital basis has been argued~\cite{Fanfarillo17}
to be quite small if the nematic order emerges as a spontaneous orbital
order (a $d-$wave Pomeranchuk instability). It can be potentially
larger, though, if the nematic order in FeSe has magnetic origin,
like in Fe-pnictides.

\section{Superconducting instabilities in the tetragonal phase}

To set the stage, we first solve the pairing problem in the tetragonal
phase, where $\Phi_{h}=\Phi_{e}=0$. To model the situation of FeSe,
our first goal is to find the region in the three-dimensional parameter
space of interactions $\left(V,W_{1},W_{2}\right)$ where the two
leading superconducting instabilities, $s^{+-}$ and $d$-wave, are
comparable. To simplify the calculations, we hereafter set $W_{1}=W_{2}$;
this does not affect our main conclusions. Denoting
the gap functions at each band $a$ by $\Delta_{a}\left(\theta\right)$, we obtain
the BCS-like non-linear gap equations in the form:
\begin{widetext}
\begin{align}
-\Delta_{h_{1}}(\theta_{1})= & T\sum_{n}\left\{ \int\frac{\rmd^{2}k_{X}}{(2\pi)^{2}}\frac{V\sin^{2}\theta_{1}+W\cos^{2}\theta_{1}}{\epsilon_{X}^{2}+\omega_{n}^{2}+|\Delta_{e_{X}}|^{2}}\Delta_{e_{X}}+\int\frac{\rmd^{2}k_{Y}}{(2\pi)^{2}}\frac{W\sin^{2}\theta_{1}+V\cos^{2}\theta_{1}}{\epsilon_{Y}^{2}+\omega_{n}^{2}+|\Delta_{e_{Y}}|^{2}}\Delta_{e_{Y}}\right\} \\
-\Delta_{h_{2}}(\theta_{2})= & T\sum_{n}\left\{ \int\frac{\rmd^{2}k_{X}}{(2\pi)^{2}}\frac{V\cos^{2}\theta_{2}+W\sin^{2}\theta_{2}}{\epsilon_{X}^{2}+\omega_{n}^{2}+|\Delta_{X}|^{2}}\Delta_{e_{X}}+\int\frac{\rmd^{2}k_{Y}}{(2\pi)^{2}}\frac{W\cos^{2}\theta_{2}+V\sin^{2}\theta_{2}}{\epsilon_{Y}^{2}+\omega_{n}^{2}+|\Delta_{e_{Y}}|^{2}}\Delta_{e_{Y}}\right\} \\
-\Delta_{e_{X}}(\fvec k_{X})= & T\sum_{n}\left\{ \int\frac{\rmd^{2}k_{1}}{(2\pi)^{2}}\frac{V\sin^{2}\theta_{1}+W\cos^{2}\theta_{1}}{\epsilon_{1}^{2}+\omega_{n}^{2}+|\Delta_{h_{1}}|^{2}}\Delta_{h_{1}}(\theta_{1})+\int\frac{\rmd^{2}k_{2}}{(2\pi)^{2}}\frac{V\cos^{2}\theta_{2}+W\sin^{2}\theta_{2}}{\epsilon_{2}^{2}+\omega_{n}^{2}+|\Delta_{h_{2}}|^{2}}\Delta_{h_{2}}(\theta_{2})\right.\nonumber \\
 & \left.\quad+W\int\frac{\rmd^{2}k_{Y}}{(2\pi)^{2}}\frac{\Delta_{e_{Y}}}{\epsilon_{Y}^{2}+\omega_{n}^{2}+|\Delta_{e_{Y}}|^{2}}\right\} \\
-\Delta_{e_{Y}}(\fvec k_{Y})= & T\sum_{n}\left\{ \int\frac{\rmd^{2}k_{1}}{(2\pi)^{2}}\frac{V\cos^{2}\theta_{1}+W\sin^{2}\theta_{1}}{\epsilon_{1}^{2}+\omega_{n}^{2}+|\Delta_{h_{1}}|^{2}}\Delta_{h_{1}}(\theta_{1})+\int\frac{\rmd^{2}k_{2}}{(2\pi)^{2}}\frac{V\sin^{2}\theta_{2}+W\cos^{2}\theta_{2}}{\epsilon_{2}^{2}+\omega_{n}^{2}+|\Delta_{h_{2}}|^{2}}\Delta_{h_{2}}(\theta_{2})\right.\nonumber \\
 & \quad+W\int\frac{\rmd^{2}k_{X}}{(2\pi)^{2}}\frac{\Delta_{e_{X}}}{\epsilon_{X}^{2}+\omega_{n}^{2}+|\Delta_{e_{X}}|^{2}}\label{Eqn:BCSGap}
\end{align}
\end{widetext}

The gaps can be parametrized as:

\begin{align}
\Delta_{h_{1}} & =\Delta_{1}\sin^{2}\theta_{1}+\Delta_{2}\cos^{2}\theta_{1}\nonumber \\
\Delta_{h_{2}} & =\Delta_{1}\cos^{2}\theta_{2}+\Delta_{2}\sin^{2}\theta_{2}\nonumber \\
\Delta_{e_{X}} & =\Delta_{X}\nonumber \\
\Delta_{e_{Y}} & =\Delta_{Y}\label{gap_functions}
\end{align}
The solutions can be decomposed into the two orthogonal channels: the
$s^{+-}$-wave state, corresponding to $\Delta_{X}=\Delta_{Y}=\Delta_{e}^{s}$
and $\Delta_{1}=\Delta_{2}=\Delta_{h}^{s}$ of opposite signs, and
the $d$-wave state, corresponding to $\Delta_{X}=-\Delta_{Y}=\Delta_{e}^{d}$
and $\Delta_{1}=-\Delta_{2}=\Delta_{h}^{d}$, leading to $\Delta_{h_{1}}=-\Delta_{h_{2}}=\Delta_{h}^{d}\cos2\theta_{h}$.

Near $T_{c}$, we can linearize the gap equations and use $T\sum_{n}\int\rmd\fvec k\frac{1}{\omega_{n}^{2}+\epsilon_{a,\fvec k}^{2}}\approx\frac{N_{a}}{2}\ln\frac{\sqrt{\Lambda\mu_{a}}}{T}$,
where $\Lambda$ is the high-energy cutoff associated with the pairing
interaction and $\mu_{a}$ is the chemical potential of band $a$.
Here, $N_{a}$ is the density of states at the Fermi level. Fixing
$W$ to be $W=0.5$ eV and solving for $T_{c}$ for varying $V$,
we find a transition from $d$-wave to $s^{+-}$ upon increasing $V$,
as shown in Fig.~\ref{Fig:SCC4}. The values of all the dispersion
parameters are listed in the Appendix A, and are consistent with those used in
our previous work~\cite{Kang18}.

\begin{figure}
\subfigure[\label{Fig:SC4:Tc}]{\includegraphics[width=0.7\columnwidth]{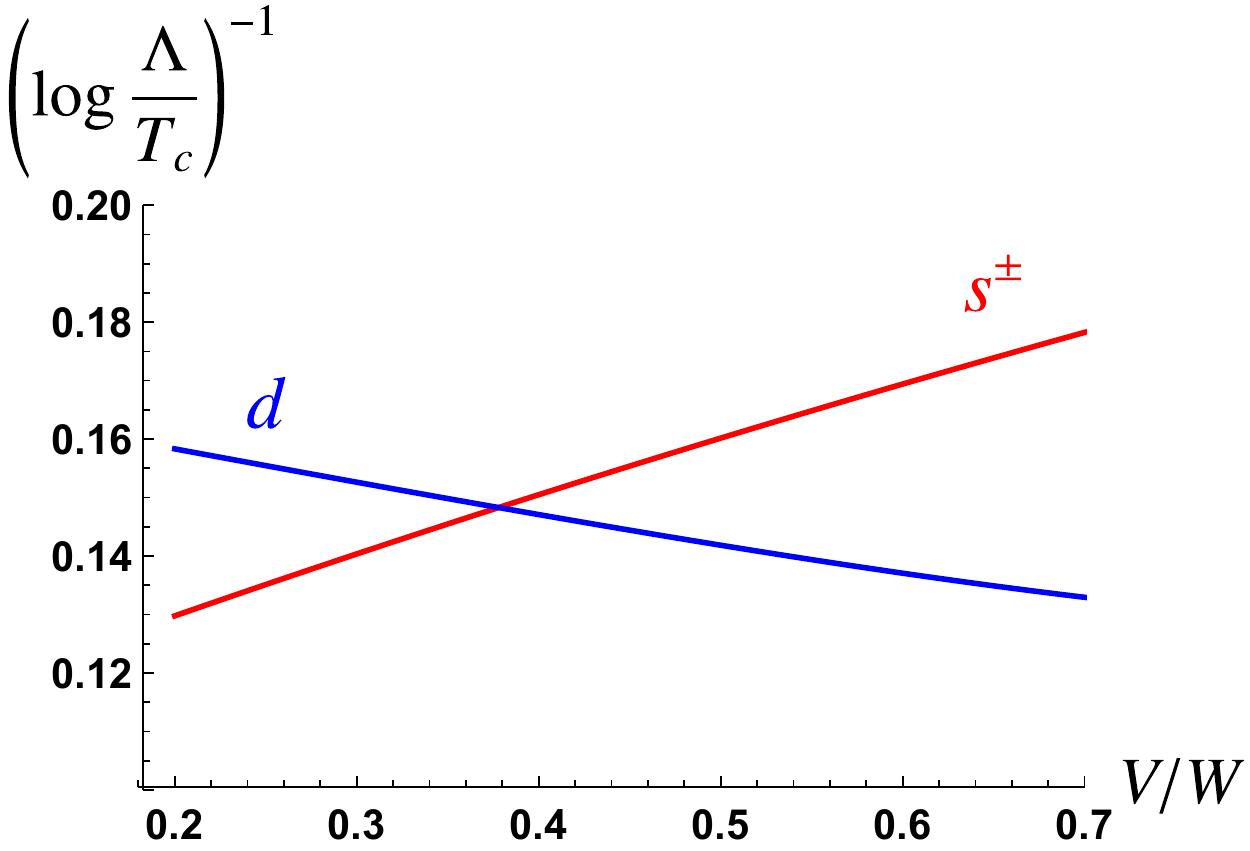}}
\subfigure[\label{Fig:SC4:Phase}]{\includegraphics[width=0.7\columnwidth]{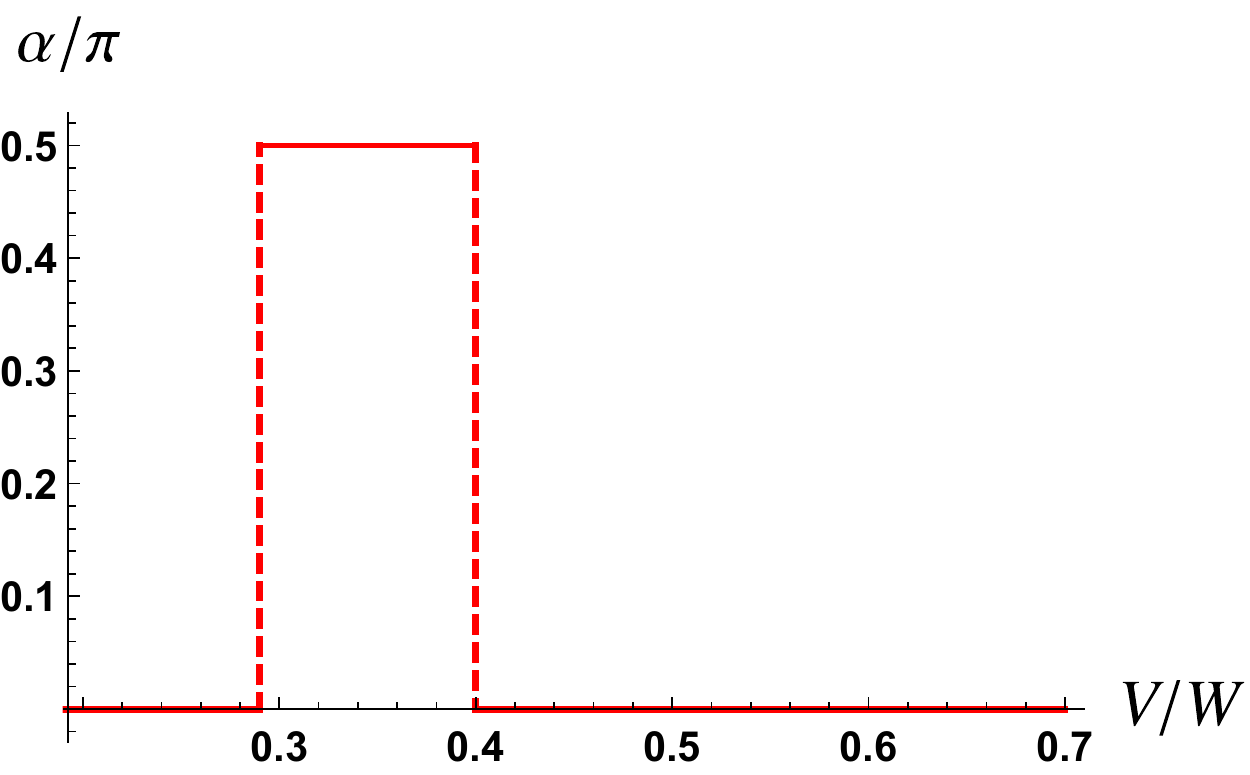}}
\caption{(a) The two leading eigenvalues $\lambda=(\ln(\Lambda/T))^{-1}$ of
the linearized BCS gap equation at $T=T_{c}$ as function of the ratio
$V/W$. Here, $V$ is the intra-orbital, inter-pocket interaction,
and $W$ is the inter-orbital, intra-pocket interaction. When $V\ll W$
, the leading pairing instability is $d$-wave. When $V\gg W$, the
leading instability is $s^{+-}$ pairing. (b) The phase difference
$\alpha$ between the $s^{+-}$-wave and $d$-wave gaps at $T=0$.
When $V\gg W$ or $V\ll W$ the pairing is either purely $s$-wave
or purely $d$-wave, and $\alpha$ is not well-defined. But when $V\sim W$,
the system spontaneously breaks time reversal symmetry by forming
an $s+id$ state with $\alpha=\pm\pi/2$.}
\label{Fig:SCC4}
\end{figure}

We also solve the gap equations at $T=0$. To search for TRSB solutions,
we introduce a relative phase between the gaps $\Delta_{X}$ and $\Delta_{Y}$,
which is related to the relative phase $\alpha$ between the
$s^{+-}$-wave and $d$-wave gaps, Eq. (\ref{1}). As shown in Fig.~\ref{Fig:SCC4},
we find that near the degeneracy point between the $s^{+-}$ and $d$-wave
states, $0.3\apprle V/W\apprle0.4$, the gap structure breaks TRS
at $T=0$, as signaled by the fact that $\alpha=\pm\pi/2$ in this
regime. Note that $\alpha$ is not well defined in the other parameter
ranges in which $s^{+-}$ and $d$-wave SC do not coexist. The resulting
schematic phase diagram at all temperatures is then that shown in
Fig.~\ref{Fig:schematic}.

\section{Superconducting instabilities in the nematic phase}
We next solve the pairing problem in the fully reconstructed nematic
Fermi surface. The onset of the nematic order, described by Eq.~(\ref{H_nem}),
has important effects on the low-energy electronic spectrum.  For the states near the hole pockets, we introduce
the Nambu operators $\Psi_{\mathbf{k}\sigma}^{\dagger}=\left(\begin{array}{cc}
d_{xz,\mathbf{k}\sigma}^{\dagger} & d_{yz,\mathbf{k}\sigma}^{\dagger}\end{array}\right)$ and write the quadratic Hamiltonian as $\bar{\mathcal{H}}_{h}=\sum_{\mathbf{k}\sigma}\Psi_{\mathbf{k}\sigma}^{\dagger}\hat{H}_{h}\left(\mathbf{k}\right)\Psi_{\mathbf{k}\sigma}^{\phantom{\dagger}}$,
with:
\[
\hat{H}_{h}\left(\mathbf{k}\right)=\hat{\tau}_{0}\varepsilon_{h_{+},\mathbf{k}}+\hat{\tau}_{1}\varepsilon_{h_{-},\mathbf{k}}\sin2\theta+\hat{\tau}_{3}\left(\Phi_{h}+\varepsilon_{h_{-},\mathbf{k}}\cos2\theta\right)
\]
Here, $\hat{\tau}_{i}$ are Pauli matrices in Nambu space and $\varepsilon_{h_{\pm}}=\left(\varepsilon_{h_{1}}\pm\varepsilon_{h_{2}}\right)/2$.
A non-zero nematic order parameter $\Phi_{h}$  splits the top
of the two hole bands and distort the hole pockets, whose new dispersions
become $E_{\pm}=\varepsilon_{h_{+}}\pm\sqrt{\Phi_{h}^{2}+\varepsilon_{h_{-}}^{2}-2\Phi_{h}\varepsilon_{h_{-}}\cos2\theta}$.
To capture the experimental result that one of the pockets is sunk
below the Fermi level \cite{Coldea17}, we set $\Phi_{h}>\mu_{h}$,
and focus only on the $E_{+}$ dispersion for the outer pocket. In terms of the original orbital operators,
the band operator for the outer pocket $h_{2\fvec k} = h_{\fvec k}$ is given still by $h_{\mathbf{k}}=i\left(-d_{xz,\mathbf{k}}\sin\theta_{\mathbf{k}}+d_{yz,\mathbf{k}}\cos\theta_{\mathbf{k}}\right)$,
but with:

\begin{align*}
\sin^{2}\theta_{\mathbf{k}} & =\half\left(1+\frac{\Phi_{h}+\epsilon_{h_{-},\fvec k}\cos2\theta}{\sqrt{\Phi_{h}^{2}+\epsilon_{h_{-}}^{2}+2\Phi_{h}\epsilon_{h_{-}}\cos2\theta}}\right)\\
\cos^{2}\theta_{\mathbf{k}} & =\half\left(1-\frac{\Phi_{h}+\epsilon_{h_{-},\fvec k}\cos2\theta}{\sqrt{\Phi_{h}^{2}+\epsilon_{h_{-}}^{2}+2\Phi_{h}\epsilon_{h_{-}}\cos2\theta}}\right)
\end{align*}
The effect of nematicity on the electron pockets is more straightforward,
as $\Phi_{e}$ simply shifts the bottom of the electron pockets centered
at $X$ and $Y$ in opposite ways, giving rise to the new dispersions
$E_{X/Y}=\varepsilon_{X/Y}\pm\Phi_{e}$. The parameters of the dispersion
are fitted with ARPES data, and listed in the Appendix A.

\begin{figure}[htbp]
\centering{}\includegraphics[width=1\columnwidth]{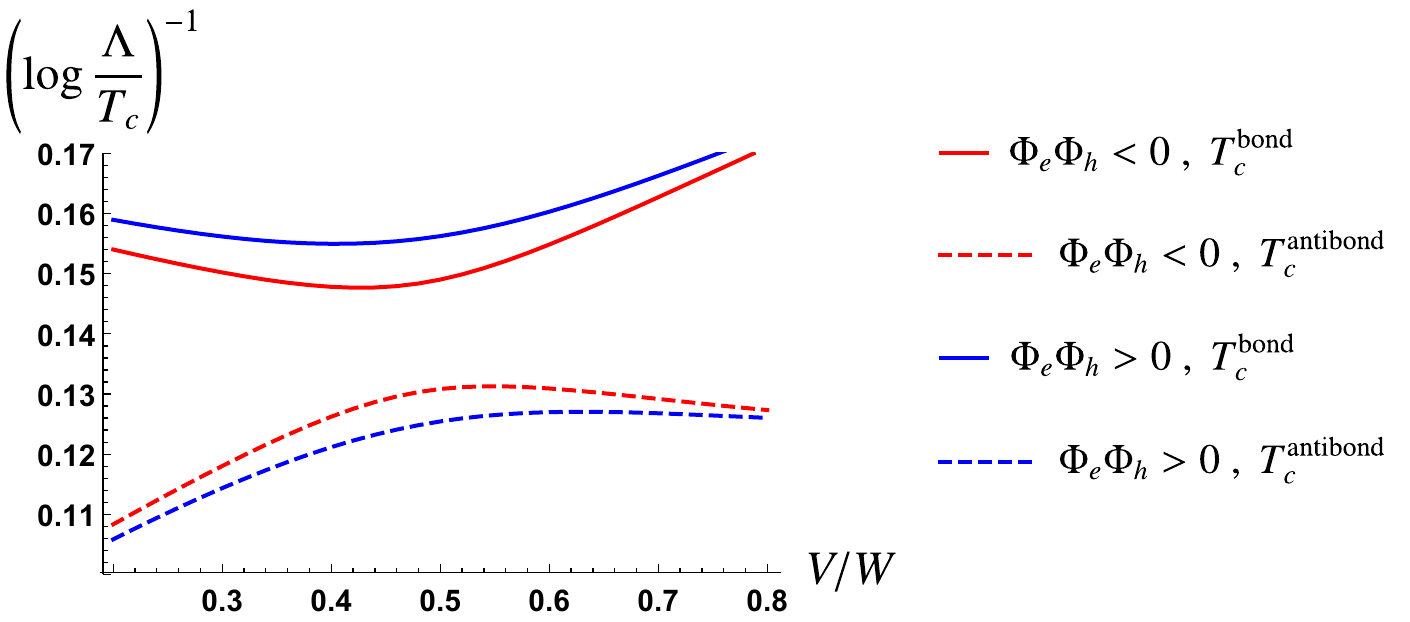} \caption{Pairing instabilities in the nematic phase. The red solid and dashed
curves refer to the sign-changing nematic state ($\mathrm{sign}(\Phi_{h})=-\mathrm{sign}(\Phi_{e})$),
whereas the blue solid and dashed curves refer to the sign-preserving
nematic state ($\mathrm{sign}(\Phi_{h})=\mathrm{sign}(\Phi_{e})$).
The splitting of the two leading pairing instabilities, corresponding
to ``bonding'' and ``anti-bonding'' mixing of the $s^{+-}$ and
$d$-wave gaps, is smaller in the case of sign-changing nematicity,
illustrating the reduced impact of nematic order on SC in this case.}
\label{Fig:NemSC}
\end{figure}

The gap equations are essentially the same as in the previous calculation,
but with $\theta\rightarrow\theta_{\mathbf{k}}$ and $\mu_{e,X/Y}\rightarrow\mu_{e}\mp\Phi_{e}$.
The solid and dashed red curves in Fig.~\ref{Fig:NemSC} shows $T_{c}$
of the two leading pairing instabilities in our model with sign-changing
nematicity, i.e. $\mathrm{sign}(\Phi_{h})=-\mathrm{sign}(\Phi_{e})$.
These two instabilities correspond to the ``bonding'' and ``anti-bonding''
mixing of the $s^{+-}$ and $d$-wave gaps, $\Delta_{s}\pm\Delta_{d}$.
Observe that the degeneracy between the $s^{+-}$ and $d$-wave gaps
is lifted by the nematic order. It is also important
to note that only the instability with a higher $T_{c}$ is realized.
Even though the second instability is not realized, the splitting
between the two solutions brings important information on how strongly
the nematic order lifts the degeneracy between the $s^{+-}$ and $d$-wave
states.

In this context, it is interesting to compare this case with the case
of same-sign nematicity, shown by the blue curves in Fig.~\ref{Fig:NemSC}.
The same parameters are used in this model except for the relative
sign of the nematic order parameters $\Phi_{h}$ and $\Phi_{e}$.
We find that the splitting between the solutions corresponding to
``bonding'' and ``anti-bonding'' mixing between the $s^{+-}$
and $d$-wave gaps is larger in the case of the same-sign nematic
order parameters.

To gain a qualitative understanding of the difference between these
two types of nematic order, we assume that the nematic order parameter
is small, and apply a Ginzburg-Landau double-expansion in terms of
the SC and nematic order parameters. Although the Ginzburg-Landau
expansion is not technically valid in the case of FeSe, where $T_{s}\gg T_{c}$,
it still provides qualitative insight for our numerical results. To
leading order, the free energy is \cite{Millis13}
\begin{align}
F(\Delta)= & a_{s}|\Delta_{s}|^{2}+a_{d}|\Delta_{d}|^{2}\nonumber \\
 & -(\beta_{h}\Phi_{h}+\beta_{e}\Phi_{e})(\Delta_{s}^{*}\Delta_{d}+\Delta_{s}\Delta_{d}^{*})+O(|\Delta|^{4})\label{Eqn:FreeEne}
\end{align}
where $a_{s}=T-T_{c}^{(s)}$ and $a_{d}=T-T_{c}^{(d)}$ refer to the
SC transition temperatures in the tetragonal phase. The last term
in the free energy shows that, in the nematic phase, the nematic order
parameters $\Phi_{e/h}$ cause a mixing between the $s$-wave and
the $d$-wave pairings. As a consequence, the single superconducting
transition at the degeneracy point is split in two, corresponding
to the ``bonding'' and ``anti-bonding'' $s\pm d$ states. The
amplitude of the splitting $\Delta T_{c}$ between the bonding and
anti-bonding mixing of the $s$-wave and $d$-wave gaps is given by:

\begin{equation}
\Delta T_{c}=\sqrt{\left(T_{c}^{(s)}-T_{c}^{(d)}\right)^{2}+\left(\beta_{h}\Phi_{h}+\beta_{e}\Phi_{e}\right)^{2}}
\end{equation}

Therefore, if the coefficients $\beta_{h}$ and $\beta_{e}$ have
the same sign, the splitting $\Delta T_{c}$ will be smaller for sign-changing
nematicity ($\Phi_{h}\Phi_{e}<0$) as compared to same-sign nematicity
($\Phi_{h}\Phi_{e}>0$). The numerical results shown in Fig. \ref{Fig:NemSC}
thus imply that the coefficients $\beta_{h}$ and $\beta_{e}$ have
the same sign. Analytical calcultions for the free energy, shown in
Appendix B, confirm this result.

\begin{figure}[htbp]
\centering \includegraphics[width=0.9\columnwidth]{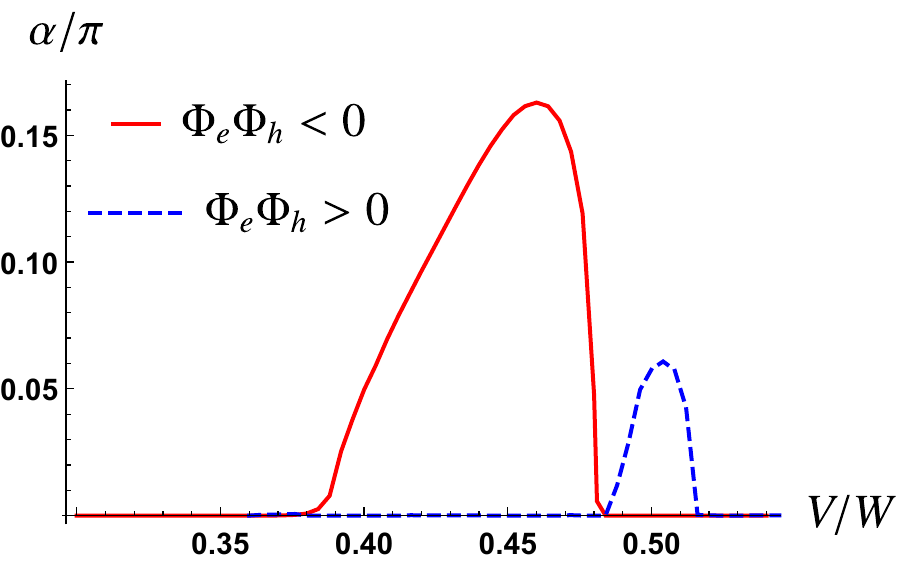} \caption{The phase difference $\alpha$ between the $s^{+-}$ and $d$-wave
gaps at $T=0$ in the nematic phase. When $\alpha$ is neither $0$,
$\pi$, or $\pi/2$ the SC state breaks both tetragonal and time-reversal
symmetries. The solid red  and dashed blue curves refer to the case of opposite-sign
nematicity and same-sign nematicity, respectively. In the former,
we find a much larger regime in which the SC state breaks time-reversal
symmetry. }
\label{Fig:NemSCPhase}
\end{figure}

The fact that $\left|\beta_{h}\Phi_{h}+\beta_{e}\Phi_{e}\right|$
is smaller for sign-changing nematicity also suggests that a TRSB
transition is more likely to take place at low temperatures in this
case as compared to the case of same-sign nematicity. To see this,
we consider higher-order terms in the free energy expansion (\ref{Eqn:FreeEne})
that are sensitive to the relative phase $\alpha$ between the $s-$wave
and $d-$wave gaps. The quartic order term is given by $\frac{\gamma}{4}(\Delta_{s}^{*}\Delta_{d}+\Delta_{s}\Delta_{d}^{*})^{2}$,
where $\gamma>0$ favors $\alpha=\pi/2$ in the tetragonal phase,
in agreement with our numerical results of the previous section. Minimization
with respect to $\alpha$ in the nematic phase leads to the solutions
$\alpha=0,\pi$, corresponding to $s\pm d$, and:

\begin{equation}
\alpha_{0}=\arccos\left(\frac{\beta_{h}\Phi_{h}+\beta_{e}\Phi_{e}}{\gamma\left|\Delta_{s}\right|\left|\Delta_{d}\right|}\right)
\end{equation}

Close to $T_{c}$, the product $\left|\Delta_{s}\right|\left|\Delta_{d}\right|$
is very small, and the $\alpha_{0}\neq0,\pi$ solution is not possible.
However, as temperature decreases and the product $\left|\Delta_{s}\right|\left|\Delta_{d}\right|$
increases, it is possible at $T^{*}<T_{c}$ for the free energy minimum
to move to $\alpha_{0}\neq0,\pi$, signaling a TRSB nematic superconducting
state (denoted here by $s+\mathrm{e}^{i\alpha}d$). Of course, smaller
$\left|\beta_{h}\Phi_{h}+\beta_{e}\Phi_{e}\right|$ leads to a higher
$T^{*}$. Thus, the regime where a TRSB nematic state is realized
is expected to be larger in the case of sign-changing nematicity as
compared to same-sign nematicity.

To go beyond this qualitative analysis, we also solved the gap equations
at $T=0$. The red curve in figure~\ref{Fig:NemSCPhase} shows the
phase difference $\alpha$ between the $s^{+-}$-wave and $d$-wave
gaps. We find $\alpha > 0$  for the range $0.39\lesssim V/W\lesssim0.48$,
signaling that the system undergoes a SC-SC transition in which time-reversal
symmetry is broken at a temperature $T^{*}$ below $T_{c}$. While
the regime with TRSB is narrower as compared to the tetragonal case,
it is enhanced by the fact that $\Phi_{e}$ and $\Phi_{h}$ have opposite
signs. Indeed, in Fig.~\ref{Fig:NemSCPhase}, the blue curve shows
$\alpha$ for the case in which $\Phi_{e}$ has the same sign as $\Phi_{h}$.
In this case, The parameter regime with TRSB SC is significantly reduced,
in agreement with our qualitative analysis.

\section{Experimental consequences: specific heat and soft mode}

The TRSB transition at $T^{*}$ belong to the Ising universality class,
and as such it is manifested in several thermodynamic quantities,
most notably as a peak in the specific heat. Because most of the entropy
related to the SC degrees of freedom is released at $T_{c}$, the
features in the specific heat at $T^{*}$ are expected to be weaker
than the jump at $T_{c}$. Interestingly, recent high-precision specific
heat measurements in FeSe reported a peak in the specific heat at
$T^{*}\approx1$ K~\cite{HHWen17}, which is consistent with a TRSB
transition.

Direct evidence for TRSB could be obtained from measurements such
as $\mu$SR and Kerr rotation, although the issues of TRSB Ising-like
domains and induced current patterns may render these measurements
challenging. We point out that recent STM data in FeSe has been interpreted
in terms of a TRSB-SC state forming at the twin boundaries \cite{Hanaguri15}.
This observation is perfectly consistent with our results, as in the
absence of nematic order, the relative phase $\alpha$ between the
$s$-wave and $d$-wave gaps becomes $\pi/2$.

Alternatively, TRSB could be detected by probing the collective modes
of FeSe. Since the TRSB transition takes place deep inside a nodeless
SC state, the electronic spectrum is fully gapped. As a result, the
SC collective modes are long-lived, as there are no quasi-particles
to promote damping. To compute the collective modes, we need to evaluate
the dynamic superconducting susceptibility. The latter can be obtained
by expanding the gap around its mean-field value $\bar{\Delta}$,
$\Delta=\bar{\Delta}+\delta$, and computing the one-loop bosonic
self-energy diagram containing the coupling between the pairing fluctuations
field $\delta$ and the fermions. In the single-band case, the bare
pairing susceptibility is a $2\times2$ matrix whose diagonal components
$\chi_{n}^{i}\left(\omega\right)$ are the normal Green's function
bubble and the off-diagonal components $\chi_{a}^{i}\left(\omega\right)$
are the anomalous Green's function bubble. They are given by:

\begin{align}
\chi_{n}^{i}(\omega) & =\frac{1}{4}\int\frac{\rmd^{2}\fvec k}{(2\pi)^{2}}\frac{|\Delta_{i}|^{2}+2\xi_{i}^{2}+\xi_{i}\omega}{\sqrt{|\Delta_{i}|^{2}+\xi_{i}^{2}}(|\Delta_{i}|^{2}+\xi_{i}^{2}-\omega^{2}/4)}\nonumber \\
\chi_{a}^{i}(\omega) & =-\frac{1}{4}\int\frac{\rmd^{2}\fvec k}{(2\pi)^{2}}\frac{|\Delta_{i}|^{2}}{\sqrt{|\Delta_{i}|^{2}+\xi_{i}^{2}}(|\Delta_{i}|^{2}+\xi_{i}^{2}-\omega^{2}/4)}
\end{align}

In the nematic superconducting state, the gaps on the three pockets
are parametrized in terms of the four gap functions $\Delta_{1}$,
$\Delta_{2}$, $\Delta_{X}$, and $\Delta_{Y}$, as discussed in Eq.
(\ref{gap_functions}). Thus, we need to introduce four pairing fluctuation
fields, resulting in an $8\times8$ bare SC susceptibility matrix
of the form:

\[
\hat{\chi}\left(\omega\right)=\left(\begin{array}{cccc}
\hat{\chi}_{n}^{h}\left(\omega\right) & 0 & \hat{\chi}_{a}^{h}\left(\omega\right) & 0\\
0 & \hat{\chi}_{n}^{e}\left(\omega\right) & 0 & \hat{\chi}_{a}^{e}\left(\omega\right)\\
\left(\hat{\chi}_{a}^{h}\right)^{\dagger}\left(-\omega\right) & 0 & \hat{\chi}_{n}^{h}\left(-\omega\right) & 0\\
0 & \left(\hat{\chi}_{a}^{e}\right)^{\dagger}\left(-\omega\right) & 0 & \hat{\chi}_{n}^{e}\left(-\omega\right)
\end{array}\right)
\]

Here, the $2\times2$ matrices are given by:

\begin{align}
\hat{\chi}_{\alpha}^{h}\left(\omega\right) & =\left(\begin{array}{cc}
\big\langle\chi_{\alpha}^{h}(\omega)\cos^{4}\varphi_{h}\big\rangle & \big\langle\chi_{\alpha}^{h}(\omega)\sin^{2}2\varphi_{h}\big\rangle\\
\big\langle\chi_{\alpha}^{h}(\omega)\sin^{2}2\varphi_{h}\big\rangle & \big\langle\chi_{\alpha}^{h}(\omega)\sin^{4}\varphi_{h}\big\rangle
\end{array}\right)\nonumber \\
\hat{\chi}_{\alpha}^{e}\left(\omega\right) & =\left(\begin{array}{cc}
\chi_{\alpha}^{e_{X}}(\omega) & 0\\
0 & \chi_{\alpha}^{e_{Y}}(\omega)
\end{array}\right)
\end{align}
with $\alpha=a,n$ and the averages are calculated with respect to
the polar angle. Within RPA, the renormalized SC pairing susceptibility
is then given by:

\begin{equation}
\left(\hat{\chi}_{R}\right)^{-1}=\left(\hat{\chi}\right)^{-1}+\hat{U}
\end{equation}
with:

\begin{equation}
\hat{U}=\left(\begin{array}{cccc}
0 & \hat{U}_{2} & 0 & 0\\
\hat{U}_{2} & \hat{U}_{1} & 0 & 0\\
0 & 0 & 0 & \hat{U}_{2}\\
0 & 0 & \hat{U}_{2} & \hat{U}_{1}
\end{array}\right)
\end{equation}
and $2\times2$ matrices:

\begin{align}
\hat{U}_{1} & =\left(\begin{array}{cc}
0 & W\\
W & 0
\end{array}\right)\nonumber \\
\hat{U}_{2} & =\left(\begin{array}{cc}
V & W \\
W & V
\end{array}\right)
\end{align}

\begin{figure}[htbp]
\centering \includegraphics[width=0.9\columnwidth]{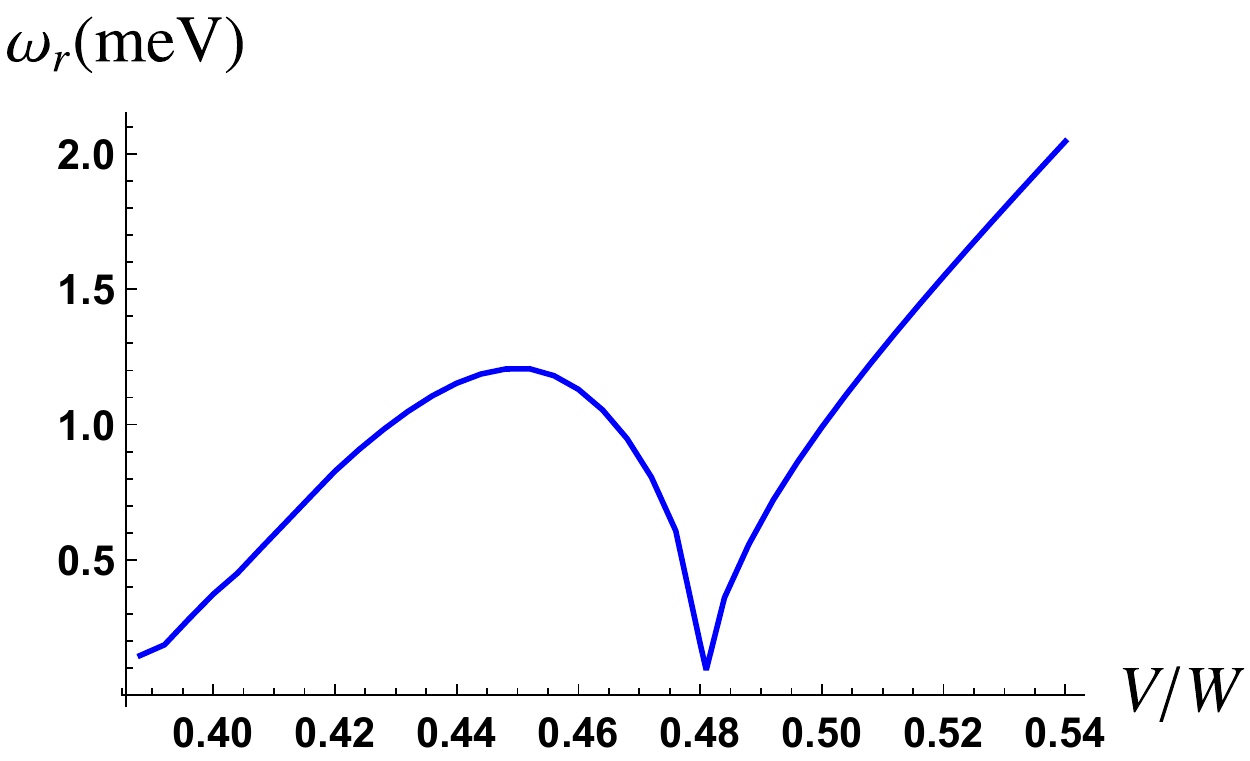} \caption{Energy $\omega_{r}$ of the collective mode associated with the relative
phase between the gaps at the $X$ and $Y$ pockets at $T=0$ as function
of the ratio $V/W$. Note that $\omega_{r}$ becomes soft when the
transition to the time-reversal symmetry-breaking state takes place.}
\label{Fig:ResFreq}
\end{figure}

Since the relative phase between $\Delta_{X}$ and $\Delta_{Y}$ assumes
a non-trivial value in the TRSB state, we expect that one of the eigenmodes
of $\hat{\chi}_{R}\left(\omega\right)$ vanishes at the transition.
Of course, because we did not consider the coupling to the density,
the mode of $\hat{\chi}_{R}\left(\omega\right)$ corresponding to
the global phase is always zero, which we ignore in our analysis,
as this mode becomes massive due to the Higgs mechanism. In Fig. \ref{Fig:ResFreq},
we plot the energy of the ``Legget-like'' mode across the TRSB transition
at $T=0$. Comparing to Fig. \ref{Fig:NemSCPhase}, it is clear that
softening occurs precisely at the boundaries delineating the regime
where the nematic SC state breaks time-reversal. Therefore, it follows
that such a soft mode should also appear at $T^{*}$. We propose Raman
experiments to verify whether such a soft mode exists in FeSe,

\section{Conclusions}

In summary, we showed that the properties of FeSe favor a second superconducting
transition at $T^{*}<T_{c}$ from a nematic $s+d$ SC state to a nematic,
time-reversal symmetry breaking $s+\mathrm{e}^{i\alpha}d$ SC state
(with $\alpha\neq0,\,\pi,\,\pi/2$). In particular, these properties
are the near degeneracy between the $s$-wave state and the $d$-wave
state; the absence of competing long-range magnetic order; and a nematic
state in which the nematic order paramater changes sign between electron
and hole pockets. We showed that this phase transition is manifested
not only in standard thermodynamic quantities, but also by softening
Legget-like mode that can be detected by Raman spectroscopy. Furthermore,
measurements such as $\mu$SR and Kerr rotation should also directly
observe time-reversal symmetry-breaking at $T^{*}$. It is tantalizing
to attribute the recently observed peak in the specific heat at $1$
K to this TRSB phase \cite{HHWen17}, although additional experiments
are necessary to elucidate the origin of this peak. Finally, we note
that the $s+d$ state that sets in below $T_{c}$ but above $T^{*}$
has been reported to be strongly anisotropic \cite{Dong16,BrianSTM17,shin,Zhou18,Borisenko18,Rhodes18}.
Time-reversal symmetry is expected to partially suppress this anisotropy,
which could also be observed experimentally.
\begin{acknowledgments}
We thank B. Andersen, L. Benfatto, S. Borisenko, D. Chowdhury, L.
Classen, F. Hardy, P. Hirschfeld, A. Kreisel, M. Eschrig, L. Rhodes,
and M. Watson for fruitful discussions. JK was supported by the National
High Magnetic Field Laboratory through NSF Grant No.~DMR-1157490
and the State of Florida. RMF and AVC were supported by the Office
of Basic Energy Sciences, U.S. Department of Energy, under awards
DE-SC0012336 (RMF) and DE-SC0014402 (AVC).
\end{acknowledgments}

\vspace{0.5cm}

\appendix

\section{Band dispersion parameters}

The band dispersion parameters used in our paper are given by:
\begin{table}[h]
\centering %
\begin{tabular}{|c| c | c | c | }
\hline
$\mu_{h}$  & $ N_1 $  & $ N_2 $  & $ \Lambda$   \tabularnewline
\hline
$13.6$meV & $0.11$eV$^{-1}$  & $0.38$eV$^{-1}$  & $1.0$eV \tabularnewline
\hline
$\mu_e$ &  $N_e$  & $\Phi_h$ & $\Phi_e$  \tabularnewline \hline
 $30$meV & $0.33$eV$^{-1}$ & $10$meV & $-18$meV  \tabularnewline  \hline
\end{tabular}
\caption{Band parameters.}
\label{TabS:Band}
\end{table}

\section{Free energy expansion}

In this Appendix, we derive the Ginzburg-Landau coefficients coupling
the nematic and superconducting order parameters:

\begin{align}
F(\Delta) & =a_{s}|\Delta_{s}|^{2}+a_{d}|\Delta_{d}|^{2}\nonumber \\
 & -(\beta_{h}\Phi_{h}+\beta_{e}\Phi_{e})(\Delta_{s}^{*}\Delta_{d}+\Delta_{s}\Delta_{d}^{*})+O(|\Delta|^{4})\label{App_F}
\end{align}

Although, as discussed in the main text, the nematic order parameter
is not necessarily small in FeSe, this expansion allows us to gain
a qualitative understanding of the differences between the cases of
sign-changing and sign-preserving nematic states. The linearized BCS
gap equations are given by (for simplicity, we set $\mu_{h}\sim\mu_{e}=\mu$):

\begin{widetext}
\begin{align}
-\Delta_{h} & =N_{e}\ln\frac{\sqrt{\Lambda\mu_{e}}}{T}\big(V\cos^{2}\theta+W\sin^{2}\theta\big)\Delta_{X}+N_{e}\ln\frac{\sqrt{\Lambda\mu_{e}}}{T}\big(W\cos^{2}\theta+V\sin^{2}\theta\big)\Delta_{Y}\nonumber \\
-\Delta_{X} & =N_{e}W\ln\frac{\sqrt{\Lambda\mu_{e}}}{T}\Delta_{Y}+N_{h}\ln\frac{\sqrt{\Lambda\mu_{h}}}{T}\left\langle \big(V\cos^{2}\theta_{h}+W\sin^{2}\theta_{h}\big)\Delta_{h}\right\rangle _{\theta}\nonumber \\
-\Delta_{Y} & =N_{e}W\ln\frac{\sqrt{\Lambda\mu_{e}}}{T}\Delta_{X}+N_{h}\ln\frac{\sqrt{\Lambda\mu_{h}}}{T}\left\langle \big(V\sin^{2}\theta_{h}+W\cos^{2}\theta_{h}\big)\Delta_{h}\right\rangle _{\theta}
\end{align}

\end{widetext}

The $s$-wave solution corresponds to $\Delta_{X}=\Delta_{Y}=\Delta_{e}^{(s)}$
and $\Delta_{h}=\Delta_{h}^{(s)}$, whereas the $d$-wave solution
gives $\Delta_{X}=-\Delta_{Y}=\Delta_{e}^{(d)}$ and $\Delta_{h}=\Delta_{h}^{(d)}\cos2\theta_{h}$.
In terms of these parametrizations, the coupled gap equations become:

\begin{equation}
\begin{pmatrix}\lambda_{s}(V+W) & 1\\
1+\lambda_{s}W & \frac{N_{h}}{N_{e}}\lambda_{s}\frac{V+W}{2}
\end{pmatrix}\begin{pmatrix}\Delta_{e}^{(s)}\\
\Delta_{h}^{(s)}
\end{pmatrix}=0
\end{equation}
and:

\begin{equation}
\begin{pmatrix}\lambda_{d}(V-W) & 1\\
1-\lambda_{d}W & \frac{N_{h}}{N_{e}}\lambda_{d}\frac{V-W}{4}
\end{pmatrix}\begin{pmatrix}\Delta_{e}^{(d)}\\
\Delta_{h}^{(d)}
\end{pmatrix}=0
\end{equation}
where we defined the coupling constants $\lambda_{\left(s,d\right)}=N_{e}\ln\frac{\sqrt{\Lambda\mu}}{T_{\left(s,d\right)}}$.
The ratios between $\Delta_{h}$ and $\Delta_{e}$ in the $s$- and
$d$-wave channels, defined as $\alpha_{\left(s,d\right)}=\Delta_{h}^{(s,d)}/\Delta_{e}^{(s,d)}$,
can be readily extracted from the equations above. We have $\alpha_{s}=-\lambda_{s}(V+W)<0$,
corresponding to an $s^{+-}$ state, and $\alpha_{d}=\lambda_{d}(W-V)>0$,
corresponding to a $d^{++}$ state (recall that the $d$-wave state
takes place only when $V<W$).

To compute the coupling constants $\beta_{e}$ and $\beta_{h}$ in
Eq. (\ref{App_F}), we first calculate the coefficients $\gamma_{e}$
and $\gamma_{h}$ defined by:

\begin{align}
\delta F_{e} & =-\gamma_{e}\Phi_{e}(|\Delta_{X}|^{2}-|\Delta_{Y}^{2}|)\nonumber \\
\delta F_{h} & =-\gamma_{h}\Phi_{h}(\Delta_{h}^{(s)^{*}}\Delta_{h}^{(d)}+c.c)
\end{align}

Straightforward calculation of the triangular Feynman diagrams gives:
\begin{align}
\gamma_{e} & =2N_{e}\frac{1-2n_{f}(\mu)}{2\mu}>0\nonumber \\
\gamma_{h} & =-N_{h}\frac{1-2n_{f}(\mu_{h})}{2\mu_{h}}<0
\end{align}

Now, using the fact that $|\Delta_{X}|^{2}-|\Delta_{Y}|^{2}=(\Delta_{e}^{(s)^{*}}\Delta_{e}^{(d)}+c.c)$,
and the results $\Delta_{h}^{(s,d)}=\alpha_{\left(s,d\right)}\Delta_{e}^{(s,d)}$
derived above, we arrive at:

\begin{equation}
\delta F=-\left(\gamma_{h}\alpha_{s}\alpha_{d}\Phi_{h}+\gamma_{e}\Phi_{e}\right)(\Delta_{e}^{(s)^{*}}\Delta_{e}^{(d)}+c.c)
\end{equation}

Therefore, we can identify $\beta_{h}=\gamma_{h}\alpha_{s}\alpha_{d}$
and $\beta_{e}=\gamma_{e}$. Since $\gamma_{e},\alpha_{d}>0$ and
$\gamma_{h},\alpha_{s}<0$, it follows that $\beta_{h},\beta_{e}>0$.
We checked that inclusion of the contributions arising from the changes
in the pairing interaction caused by nematicity does not alter this
result.
\end{document}